\documentclass[manuscript,screen]{acmart} % Comment out for submission
\AtBeginDocument{%
  }

\copyrightyear{2024}
\acmYear{2024}
\setcopyright{acmlicensed}\acmConference[FAccT '24]{The 2024 ACM Conference on Fairness, Accountability, and Transparency}{June 3--6, 2024}{Rio de Janeiro, Brazil}
\acmBooktitle{The 2024 ACM Conference on Fairness, Accountability, and Transparency (FAccT '24), June 3--6, 2024, Rio de Janeiro, Brazil}
\acmDOI{10.1145/3630106.3658898}
\acmISBN{979-8-4007-0450-5/24/06}

\begin{document}

%% Title and short title
\title[AI Art is Theft]{AI Art is Theft: Labour, Extraction, and Exploitation}
\subtitle{Or, On the Dangers of Stochastic Pollocks}

%% Author information
\author{Trystan S. Goetze}
\email{tsgoetze@cornell.edu}
\orcid{0000-0002-3435-3264}
\affiliation{%
  \institution{Sue G. and Harry E. Bovay Program in the History \& Ethics of Professional Engineering, Cornell University}
  \streetaddress{136 Hoy Road}
  \city{Ithaca}
  \state{New York}
  \country{USA}
  \postcode{14853}
}

%% Abstract
\begin{abstract}
Since the launch of applications such as \textsc{dall•e}, Midjourney, and Stable Diffusion, generative artificial intelligence has been controversial as a tool for creating artwork. Some writers have presented worries about these technologies as harbingers of fully automated futures to come, but more pressing is the impact of generative AI on creative labour in the present. Already, business leaders have begun replacing human artistic labour with AI-generated images. In response, the artistic community has launched a protest movement, which argues that AI image generation is a kind of theft. This paper analyzes, substantiates, and critiques these arguments, concluding that AI image generators involve an unethical kind of labour theft. If correct, many other AI applications also rely upon theft.
\end{abstract}

\ccsdesc[500]{Social and professional topics~Intellectual property}
\ccsdesc[300]{Social and professional topics~Codes of ethics}
\ccsdesc[500]{Social and professional topics~Automation}
\ccsdesc[500]{Social and professional topics~Socio-technical systems}
\ccsdesc[300]{Social and professional topics~Economic impact}
\ccsdesc[300]{Social and professional topics~Employment issues}
\ccsdesc[300]{Computing methodologies~Machine learning}

%% Keywords
\keywords{generative AI, computer art, diffusion models, text-to-image AI, philosophy, computer ethics, AI ethics, data colonialism, John Locke, intellectual property, labour, automation}

%% Teaser image
% \begin{teaserfigure}
%   \includegraphics[width=\textwidth]{Theatre-Dopera-Spatial-Stolen.jpg}
%   \caption{Artists contend that AI-generated images, such as the above, are created using stolen art.}
%   \Description{Jason Allen's ``Theatre D'Opéra Spatial,'' an AI-generated image of three figures in a dark room, looking out on a circular portal to a brightly lit landscape. The image is bisected diagonally by a yellow strip on which is written "CONTAINS STOLEN PROPERTY" in heavy black text.}
%   \label{fig:teaser}
% \end{teaserfigure}

%%
%% Main body
%%
\maketitle

\emph{This is an author-produced, post-review preprint of an article accepted for publication in \emph{FAccT'24: Proceedings of the 2024 ACM Conference on Fairness, Accountability, and Transparency}. Please cite the version of record, which will be available at: <doi.org/10.1145/3630106.3658898>.}

\section{Introduction: A Contest, Protests, and Feminist Pragmatist Computer Ethics}
In August 2022, a digital image titled \textit{Théâtre D’opéra Spatial} won first place in a Colorado State Fair fine arts competition. The eerie image depicts three human-like figures in a dark hall, looking out through a wide circular window on a brightly lit landscape. However, the person who submitted the image for the competition, Jason Allen, did not create the image entirely by himself. Rather, much of the image was generated using an artificial intelligence application called Midjourney, with only some touch-up work done by Allen. Similar to applications such as \textsc{dall•e} or Stable Diffusion, Midjourney is an example of text-to-image generative AI, a computer application that uses machine learning to produce images in response to user-provided text prompts.

It isn’t clear whether the judges of the competition or Allen’s competitors fully understood the nature of Midjourney at the time of Allen’s win. As such, the initial conversation in news and social media concentrated on the question of whether Allen’s use of AI amounted to \emph{cheating} \cite{Gault22}. One could make an analogy with sports: is using text-to-image AI in a fine art competition more like the advantage of a reduced-drag swimsuit \cite[cf.][]{Barrow12} or more like taking a bus to the finish line instead of running a marathon \cite[cf.][]{Subramanian11}?

Another question we could ask is whether the outputs of text-to-image AI are \emph{art} properly so-called. Since the human contribution to the creation of these images consists primarily in the initial text prompt, it is tempting to argue that the creativity involved is so minimal that it would be more appropriate to think of the outputs of text-to-image AI as merely the result of a mechanical process, with no genuine artistic skill on display. Some online art sharing communities have used such arguments to justify banning AI-generated content \cite[e.g.][]{baio_online_2022,cole_i_2023}.

Let's call these two lines of reasoning \emph{delegitimizing arguments} about AI-generated images; for they both try to show that it is a category error to call these images ``art.'' Whatever philosophical interest delegitimizing arguments may have, we should be wary of them. New and disruptive art forms often give rise to delegitimizing arguments from the artistic establishment, but are swiftly dismissed as tastes change. This pattern repeats from impressionist painting\footnote{For example, art critic Louis Leroy is widely quoted as having written, in a review of an impressionist exhibition, that ``Wallpaper in its embryonic state is more finished'' than Claude Monet’s paintings.} to the selfie.\footnote{After Oxford Dictionaries chose ``selfie'' as the word of the year for 2013, several media outlets ran pieces decrying how the trend of taking pictures of oneself and posting them to social media was a sign of narcissism and a decline of culture in the West \cite{nunberg_narcissistic_2023}.} In fact, nearly identical delegitimizing arguments were made about photography in the nineteenth century by painters and art critics. For example, here is what one anonymous critic wrote in 1865:
\begin{quote}
Photography has reached such perfection of late, that evident confusion has arisen in the minds of many persons respecting the relative difference between it and Fine Art... [A]rt differs from any mechanical process in being ``the expression of man’s delight in God’s work''... All labor of love must have something beyond mere mechanism at the bottom of it. \cite{anon_art_1865}
\end{quote}
According to this critic, because photographs are produced by a rapid mechanical process instead of the creative labour of a painter, they are not truly fine art, if art they be at all. While today we might disagree in the case of photography, this argument might be tempting to apply to text-to-image AI.\footnote{It is worth noting that art created with the assistance of computers was subject to delegitimizing arguments well before the advent of text-to-image AI. See, for example, early criticism of the work of Vera Molnar, who began using computer algorithms in her artistic practice in the 1960s \cite{williams_vera_2023}.}

There is another point of comparison with the onset of consumer photography: the rapidity with which photography could produce images from real life posed a threat to the art world of the day. As critic Charles Baudelaire wrote in 1859, ``[photography], by invading the territories of art, has become art’s most mortal enemy'' \cite[][p. 230]{baudelaire_modern_1955}. Of course, today, there are very few professional painters, while nearly all of us have become amateur photographers thanks to smartphones and other camera-equipped devices that we carry with us. As was feared by visual artists two centuries ago, paintings are now a luxury product, having been largely displaced by photos.

There is a similar fear that human-made visual art of \emph{all} genres may similarly be replaced by text-to-image AI. Thus, we might view delegitimizing arguments as a kind of motivated reasoning---maybe text-to-image AI \emph{can} produce genuine art, and these arguments are driven by economic desperation rather than substance.\footnote{Thanks to David Collins for discussion of this point.} Some human jobs may be replaced, but perhaps it will be like other cases of automation, where menial tasks were taken over by machines, and humans could get on with more valuable work.\footnote{These are the cases for automation typically emphasized by thought leaders in business \cite{mcafee_machine_2017}.}

However, unlike the dull, dirty, and dangerous tasks that were once the targets of automation, which are arguably unfair for anyone to be required to do in order to survive, what text-to-image AI promises to replace is the human practice of visual art itself. Indeed, creative businesses have already started replacing human artists with AI in game development, film production, interior design, and advertising \cite{edwards_netflix_2023, sharf_marvel_2023, zhou_ai_2023, roose_AItransforming_2022}. Co-founder of DreamWorks Animation Jeffrey Katzenberg has claimed that AI might soon replace up to 90\% of all jobs in animation \cite{maglio_jeffrey_2023}. Jason Allen himself quipped in an interview that ``This isn’t going to stop... Art is dead, dude. It’s over. AI won. Humans lost'' \cite{roose_AIprize_2022}.

In fact, philosopher John Danaher remarks that the threat posed by the current wave of automation to valuable pursuits is distressingly general:
\begin{quote}
[I]t is difficult to contain the rise of automating technologies in a way that ensures that it only displaces forms of work that are ill-suited to human flourishing and meaning... If we think, broadly, about the domains of activity that are most commonly associated with flourishing and meaning---the Good, the True, and the Beautiful---we already see evidence for the encroachment of automation. \cite[][p. 104--5]{danaher_automation_2019}
\end{quote}
Danaher worries, more broadly than visual art, that we are on the cusp of automating a wide range of activities that are necessary for humans to live flourishing lives. These include moral and political reasoning, scientific and scholarly discovery, and artistic and cultural creation.\footnote{Sam Altman, the CEO of OpenAI (the company that created \textsc{dall•e} and ChatGPT) frequently claims that something like an all-purpose general AI assistant that can replace nearly all human labour, including these intrinsically valuable activities, is his ultimate goal \cite[e.g.,][]{altman_soon_2022}} If engagement in these activities is needed for human beings to live well, automating them away would be radically bad for us, felling the tree of human value at the roots.

In response to these threats to human value, Danaher argues that we must prepare to transition to a radically new structure of society, where either human beings merge with computing technology or retreat to a virtual world. But these longterm proposals leapfrog over what we might do in the present to ensure that flourishing human lives remain possible \emph{without} needing to redefine what it means to be human. No technology is inevitable, and when some of the most powerful actors in society---such as multibillion- or trillion-dollar companies---push the adoption of technologies known to be harmful, collectively we must resist.\footnote{Compare the approach to analysis of the social impacts of AI in the academic and outreach work of Emily Bender, Timnit Gebru, and their colleagues \cite{bender_dangers_2021, dair_institute_mystery_nodate}}

For these reasons, it is important to centre a different perspective when filling the gaps in our norms around the new possibilities of AI-generated images.\footnote{Here I follow James Moor: one of the purposes of philosophical reflection on computer ethics is the filling of what he calls \emph{policy vacuums} created by the new possibilities for action enabled by emerging technologies \cite{moor_what_1985}.} As Nancy McHugh has argued, when considering impacts of science and technology on society, we must refer not only to mainstream perspectives from within science and technology research, but also to the perspectives of marginalized communities who are resisting the negative impacts of science and technology \cite{mchugh_limits_2015}.\footnote{McHugh uses this approach to analyze various cases, such as the ongoing health effects of chemical weapons used by the USA in Viet Nam. She criticizes public health policy for inadequate action due to over-reliance on scientific studies that fail to capture the actual lived experience of Viet people.} Drawing on epistemological theory developed by feminist philosophers and the American pragmatist philosopher John Dewey,\footnote{For the feminist epistemological tradition, see, among others, \cite{haraway_situated_1988, harding_science_1986, code_what_1991, collins_black_2000, sandoval_methodology_2000}. The most relevant components of Dewey’s work include \cite{dewey_quest_1984} and \cite{dewey_ethics_1985}.} McHugh demonstrates that marginalized communities have epistemic and moral authority that should inform public decision-making regarding the impacts of science and technology.

In this paper, I apply this feminist pragmatist approach by centring the experience and knowledge of artist-led protests against text-to-image AI.\footnote{See \cite{benjamin_fuckthealgorithm_2022} for another recent article centring the moral epistemology of a protest movement directed at the harms caused by algorithmic systems. For another feminist pragmatist project focused on labour and social movements, see \cite{anderson_quest_2014,anderson_moral_2015,anderson_social_2014,anderson_social_2015}.} Since Jason Allen’s prize win, the artistic community has become increasingly aware of generative AI applications, and has organized a social movement against them. Their tactics include flooding art sharing platforms with anti-AI messaging, call-outs of media that employ AI-generated images instead of human-produced work, banning or flagging AI-generated art on art-sharing sites, image processing software that confuses or damages the AI training process, and legal action against AI developers \cite{plunkett_artists_2022,baio_online_2022,weatherbed_artstation_2022,hill_this_2023,veltman_new_2023,noauthor_create_nodate}. Since any well-organized protest movement needs a shared understanding of what they are protesting against,\footnote{Otherwise, their cause will be vague, their messaging inconsistent, and their actions ineffectual. Consider the failure of Occupy Wall Street \cite{ehrenberg_what_2017}.} despite not being traditional experts in AI development, artist-protestors have educated themselves about these applications. They have also developed a sophisticated understanding of their distinctive experiences of the effects of these technologies. Moreover, because of the structure of the global economy for artistic creations, the worst effects of the proliferation of text-to-image AI are being felt first by artists who are already socially marginalized on various axes of oppression. Much animation and game production, for example, is done by low-paid workers in East Asia. Since these industries are structured to find the cheapest possible labour, artists dependent on this kind of work are doubly vulnerable due to their marginalization within the global economy. Given the framework outlined above, artist-protestors’ accounts of text-to-image AI are worth centring as we seek to understand the effects of these technologies, the possibilities they present, and how our norms should change in response to their proliferation.

Perhaps the central claim of the artistic community’s anti-AI protest is that the most common AI image generation applications involve a kind of \emph{theft}. In order to set up the argument, I first provide a rough-and-ready non-technical explanation of how the most popular AI image generation applications work (§2). I then analyze the concept of theft and comment on specific claims made by artist-protestors (§3). While multiple senses of theft are relevant to text-to-image AI, the central sense in protest arguments emerges as theft of creative \emph{labour}. I substantiate the argument that text-to-image AI steals artists’ creative labour by drawing on John Locke’s account of the connection between labour and property rights. This argument raises questions, however, about how to distinguish the ways in which human artists draw on one another’s works from the way in which AI models do so. I answer this concern by showing that text-to-image AI differs from existing practices in the artistic community by introducing distributive injustices, systematically violating consent and related norms of respect, and repeating a colonialist pattern of extraction and exploitation (§4). I conclude by outlining some implications of the account for AI development more generally (§5).

\section{AI Image Generation}
There are multiple kinds of AI image generation. Early approaches, such as those of Vera Molnar and Harold Cohen, involved hand-coded algorithms and expert systems \cite{williams_vera_2023,garcia_harold_2016}. Another technique uses generative adversarial networks (GANs) \cite{karras_stylebased_2019}. However, the most recent, popular, and controversial wave of AI image generators, including Midjourney and \textsc{dall•e}, is based on diffusion models; I focus on these systems for the rest of the paper. (References to ``text-to-image AI'' and ``AI image generators'' should be taken to refer to applications of diffusion models unless otherwise noted.)\footnote{This focus is meant to expose the ethical problems particular to diffusion models, which are the most popular publicly accessible AI image generation applications. However, it is worth noting that discourse around AI image generation all too often runs together these distinct families of AI applications, when in fact their technical differences can give rise to ethical differences. Expert systems and hand-coded generative AI are much less likely to be vulnerable to the arguments that AI art is theft since they do not rely on the data collection techniques at issue. Whether GANs are vulnerable to these arguments will depend on the size and source of the training dataset: a GAN trained on a smaller dataset containing materials in the public domain or obtained with the permission of the original creators would avoid the ethical faults of diffusion models. Thanks to Jane Adams for discussion of this point.} The following is a simplified and non-technical overview that preserves the philosophically relevant details.\footnote{For additional technical details, see \cite{ramesh_hierarchical_2022,ramesh_zero-shot_2021}.}

The AI involved in text-to-image diffusion models is not ``intelligent'' in the way that human beings or other animals are intelligent. It has no internal experience, no desires, no autonomy, no embodiment, and no creativity. The term \emph{artificial intelligence} is used here, as in most contemporary contexts, to refer to a complicated machine learning model that replicates results usually only possible through the actions of intelligent beings. To echo the Victorian critic of photography, it is ``mere mechanism.'' Still, the results can be impressive enough to give the illusion of creative thought.

To create AI of this kind, the first step is to amass a large set of training data. In the case of text-to-image AI, the training data consist of a very large number of images with accompanying text descriptions; one commonly used dataset, LAION-5B, contains nearly 5 billion image-annotation pairs \cite{schuhmann_laion-5b_2022}.\footnote{During the writing of this paper, LAION-5B was temporarily unavailable after multiple researchers reported that it contained a substantial amount of child sex abuse material, despite LAION's claims to vet their dataset for safety \cite{birhane_into_2023,thiel_identifying_2023}. It isn't clear how many AI models still in operation were trained on versions of LAION-5B containing this material.} These image-annotation pairs are collected using web crawlers and web scrapers---applications that browse the public Web, indexing or downloading data as they go. These image-annotation pairs are used to train machine learning algorithms, producing a model that can predict the text that is likely to be associated with a given image. By finding a point in the ``latent space'' of the model (areas of the model which are between items from the training data) that corresponds to properties in a supplied image, a text annotation can be derived. The original use case of these models was automatic generation of image descriptions \cite{radford_learning_2021}.

What text-to-image AI does next is essentially to reverse how the model is used. Instead of providing an image to be annotated, the user inputs a textual description of the image that they desire. The system then generates a representation of an image in its mathematical model that corresponds to that text prompt. To turn that representation into an actual image, the system takes an image of random noise, and applies techniques developed to remove noise from images, using the representation of the desired image as a basis. In other words, the system treats a picture of noise as if it were a heavily damaged or compressed version of the desired image, and gradually ``clears away'' the noise to create an image.

With this sketch of text-to-image AI in place, let's consider arguments made by artists protesting against it.

\section{Theft}
The central claim made by artist-protestors is that generative AI is or relies upon a kind of art theft. There are, however, several different senses of the concept of \emph{art theft}, and language used by protestors frequently seems to invoke several at once. Clarifying the conceptions of theft at issue, and whether one is central to protest arguments, is thus essential to understand the case against text-to-image AI. As we will see, these clarifications also serve to rebut counterarguments which are directed at senses of theft which turn out to be peripheral. In this section, I distinguish three conceptions of art theft, and relate each of them to arguments made by protestors.

\subsection{Heist}
Perhaps the prototypical sense of art theft is that of an \emph{art heist}, which we may define as the removal of artworks from their proper place without authorization. The wrong of art heist is straightforward: the thief has deprived the victim of their property, which is (generally) highly valuable to the rightful owner---monetarily, aesthetically, or otherwise---and may have cultural value to society as well. Those kinds of value of course cannot be realized by the proper owner of the stolen material because of its unauthorized removal, hence the harm of the heist.

Some protest messaging around AI image generators uses the term ``heist'' and other phrases suggesting this sense of theft. For example, here is an excerpt from an open letter published by a group of artists, journalists, and academics, which urges publishers and media organizations to eschew AI-generated images in articles and books:
\begin{quote}
AI-art generators are trained on enormous datasets, containing millions upon millions of copyrighted images, harvested without their creators’ knowledge, let alone compensation or consent. \emph{This is effectively the greatest art heist in history}. Perpetrated by respectable-seeming corporate entities backed by Silicon Valley venture capital. \emph{It’s daylight robbery.} \cite[][emphases added]{center_for_artistic_inquiry_and_reporting_restrict_2023}
\end{quote}
Taken literally, this is obviously false. AI image generators are trained on \emph{digital} images. Even granting that some of those images are representations of physical artworks, nothing has been physically taken. In a sense, the images have been taken without authorization through the data scraping process, but because of their digital nature, no artist or other party has been deprived of the original artwork, so the harms associated with a heist are not instantiated. The heist sense of art theft is thus being invoked metonymically, for emphasis, in protest texts like the above---and any counterarguments aimed at the language invoking physical forms of theft must be taken to be in bad faith. Some other sense(s) of theft must be more central to the ethical claim being made.

\subsection{Plagiarism}
A common claim against text-to-image AI is that these applications \emph{plagiarize} the works of human artists. To plagiarize a creative work is to illicitly claim creative ownership over it, that is, to assert that one is the creator of it, when in fact the work was---in whole or in significant part---created by someone else.\footnote{The addition of ``illicitly'' is needed to avoid capturing cases, such as ghost writing, where the creator has, by mutual agreement, ceded to another their right to be identified as such, as well as responsibility for the contents of the work.} The wrong of plagiarism is distinct from that of a heist: the creator of the work may still possess the original, so they have not lost the work itself. Rather, what the plagiarist attempts to steal is the \emph{credit} and associated rewards for having made the work---plagiarism is \emph{theft of responsibility}. Plagiarism has several subspecies, of which I will discuss three: copying, delegation, and style theft.\footnote{Other potential senses of plagiarism include remixing and failure to attribute source material. \emph{Remix} art forms, including collage and audio sampling, have been controversial in some artistic communities, while being embraced by others, precisely because they reuse portions of existing works without authorization \cite{mcleod_cutting_2011,del_peral_using_1989,passero_copyright_1995}. And a central argument in a high-profile lawsuit is that text-to-image AI steals existing works through a collage-like process \cite{andersenVstability}. However, this argument concerns a later stage of the text-to-image AI process than I am interested in---this paper is focused on the training stage more than the generation stage. But it strikes me that, if successful, this argument would prove too much, making physical collage and similar art forms impermissible. \emph{Failure to attribute source material} is a common form of plagiarism in academic contexts. Typically, it results from a student copy-pasting text from a book or article and failing to provide a citation, implicitly claiming responsibility for words that are not their own. Sometimes, artistic communities ban AI-generated images on the grounds that source material is not attributed; one example is the \texttt{r/Worldbuilding} subreddit \cite{reddit_rworldbuilding_2023}. However, I find this argument odd given that artists are not usually required to produce a list of sources for their work in other contexts. I come back to norms of art-sharing in §4.} 

\subsubsection{Copying}
The most obvious sense of plagiarism is that of copying, where the plagiarist claims responsibility for a specific pre-existing creative work, in whole or in part. At first blush, we might expect text-to-image AI not to produce copies, except when explicitly commanded to do so. As mentioned, when systems such as \textsc{dall•e} generate an image from the representation extracted from the model’s latent space, they start from a field of random noise, which is repeatedly ``enhanced'' to bring out the desired image. We might think that the injection of randomness into the process should ensure that the system’s outputs are always unique, unless the same random seed is re-used.

However, AI image generators can still produce copies even when not told to do so. A group of computer scientists demonstrated that models built on the techniques described in §2 sometimes ``memorize'' images in their training data \cite{carlini_extracting_2023}. This means that for some text prompts, the system will produce an image that is essentially a low-resolution duplicate of an image in the training set. With regard to plagiarism, memorization means that a user of an AI image generator could---intentionally or accidentally---produce a copy of an image in the training set, and claim it as their own creation, thus engaging in plagiarism. However, these instances are uncommon overall, so this cannot be the central sense in which text-to-image AI is theft.

\subsubsection{Delegation}
Another form of plagiarism is an illicit delegation of work. As any educator knows, there is always a risk that a student may find someone who will agree to do their assigned work for them. Whether there is an exchange of money or not, this arrangement is treated as plagiarism despite the fact that it does not involve any unauthorized copying, because this is still a form of theft of responsibility. If successful, the student gets the academic credit for completing the assignment, even though they did not.

Generative AI in all genres has prompted renewed reflection on how to deal with this kind of plagiarism in education. Much of the focus has been on ChatGPT and similar text generators,\footnote{A cottage industry to which I have no shame in having contributed \cite{goetze_chatgpt_2023}.} but we can just as easily apply these worries to visual art. A student in a digital art class using an AI image generator to create their submission for a digital painting assignment would be plagiarizing, as the process of creation is not properly attributable to the student. At best, the student would be responsible for only a part of the process, namely, the text prompt they entered.

On the other hand, suppose the assignment is instead to write an effective prompt for the AI, and to touch up the result by hand. In this case, the use of AI is disclosed to all parties and the credit is attributed for the writing of an effective prompt, as well as the selection and editing of the outputs. This kind of case isn’t plagiarism because there is no misattributed responsibility.

So, whether AI plagiarizes or facilitates plagiarism, in the sense of delegation, can be contextual. Moreover, the potential wrongs of delegation are those of deception, not theft: the illicit activity involved is the obfuscation of the true source of the image produced, not the use of the AI image generator \emph{per se}. Delegation is thus not a good candidate for the needed conception of theft.

\subsubsection{Style Theft}
The last kind of plagiarism I will discuss is theft of another artist's distinctive style, or of a design that they created. Style and design theft is familiar in the technology sector---e.g. Apple's 2011 lawsuit alleging that Samsung copied the ``look and feel'' of the iPhone \cite{appleinsider_staff_apple_2011}. In artistic communities, there is a fuzzy line between taking inspiration from another artist's style and ripping it off, as captured in this piece of advice from author Austin Kleon:
\begin{quote}
First, you have to figure out who to copy. Second, you have to figure out what to copy. Who to copy is easy. You copy your heroes---the people you love, the people you're inspired by, the people you want to be... What to copy is a little bit trickier. Don't just steal the style, steal the thinking behind the style. You don't want to look like your heroes, you want to see like your heroes. \cite[p. 36]{kleon_steal_2012}
\end{quote}
Drawing on wisdom from artists in many different media, Kleon is capturing both an element of how one learns a creative craft---take inspiration from those who have come before---and an artistic norm---when what you produce is too much like the work of another artist, you are plagiarizing. Under the definition of plagiarism I am using, the style thief is effectively taking credit for a style or design that was developed by someone else.

Text-to-image AI doesn’t just enable style theft: the ability to imitate the styles of specific artists is a popular feature of these applications. For example, shortly after Stable Diffusion was released, its users discovered that they could produce high-quality fantasy scenes by adding the name of a Polish artist, Greg Rutkowski, to their text prompts \cite{heikkila_this_2022}. Stability AI eventually disabled this function with regard to Rutkowski’s work, but only after he publicly objected.

Rutkowski and other artists claim that they have seen a drop in interest in their work-for-hire as AI image generators gained the ability to make convincing imitations of their style. Lost opportunity is also an allegation of harm in a lawsuit filed by three artists against Stability AI: ``The harm to artists is not hypothetical---works generated by AI Image Products `in the style' of a particular artist are already sold on the internet, siphoning commissions from the artists themselves'' \cite{andersenVstability}. The harm of lost opportunity is precisely what we might expect from a form of theft of responsibility: clients are driven away from the original creator and towards cheap imitators.

Style theft can also produce wrongs associated with other forms of deception, such as reputational damage. For example, one of the plaintiffs in the case just cited, Sarah Andersen, is the creator of a popular webcomic with a distinctive black-and-white style, simple and expressive cartooning, and a recognizable central character. Andersen’s popularity, combined with misogynist online culture, have led to her being targeted for harassment in the past. In one event, alt-right trolls edited her work to contain racist messages, and she had to clarify to concerned and angry readers that the edited comics were not her own creations. Soon afterward, Andersen learned that AI image generators had obtained the ability to mimic aspects of her style \cite{andersen_ai_2022}. While the results at the time were crude, it is easy to imagine bad actors using a future version of these applications to scale up their harassment campaign.

Then again, as Kleon's advice implies, the distinction between inspiration and style theft is subtle. A defender of text-to-image AI might argue that these applications and their users are engaged in processes similar to a novice artist who imitates the work of another as part of honing their craft. They could accept that an overreliance on style imitation through AI is problematic, but is simply a stepping-stone towards more original prompt engineering, just as a human artist imitating the style of an artist they admire is a step towards developing their own. The wrongs associated with style theft, while rampant, would not yet establish that any use of text-to-image AI is \emph{theft}. At most, the objector might conclude, they illustrate immature artistic practices and problematic behaviour common on the web, not something distinctive to AI art.

\subsection{Labour Theft}
The last sense of art theft that I will discuss is \emph{labour theft}. Theft of labour is a familiar concept in Marxist thought, as Karl Marx’s critique of capitalism is rooted in an argument that this system structures economic relations such that workers are rewarded less than the full value of their labour, which enables profit for the capitalist \cite{marx_capital_1976}. The kind of labour theft at issue in the creation of AI image generators is different, however, as the structure of the labour relations is distinct. On Marx’s account, worker exploitation arises because capitalists control the means of production—resources, land, and logistical networks—while labourers have only their labour to sell. But in the case of AI image generators, artists aren’t selling their labour to AI developers at all. Furthermore, artists themselves own (or have purchased licences to use) the means of production, such as art supplies and software applications. If AI developers are stealing creative labour, they are doing so in a different way.

We can get a clearer idea of how artists conceive of creative labour theft in this case by examining arguments made by anti–AI art protestors themselves. Here is a lengthy quote from a since-deleted tweet by a digital artist:
\begin{quote}
This is the argument AI evangelists are trying to make. Bread is just bread, you make bread and I make bread, so whoever’s bread is better people will buy. Seems fair, right?

\hspace*{3mm}But to make bread, you require resources. If you’re a baker and you want to make and sell bread, you need to buy your wheat... You develop relationships with people who provide you with these resources, compensate them for those resources, make your bread, and sell it... 

\hspace*{3mm}What’s happening with AI is that people have built robots who will fly to the farmer’s field and harvest their wheat without consent and deliver it to bakers. The bread is still bread, but bread cannot exist without wheat, and the wheat was stolen from the people who make the wheat and are trying to pay their rent. AI evangelists are arguing ``Bread is bread, why are you mad?'' while artists are arguing, ``I’m the one who grew the wheat, and you stole a portion of my income without consent.''

\hspace*{3mm}Bread cannot be made without wheat. AI cannot be made without training data… a product cannot exist without input and the input is being robbed on an industrial scale without any concern for consent, ethics, or regulation. \cite{andantonius_silver_2022}
\end{quote}
This argument makes clear that the central claim is one of labour theft, but in a way that is different from a Marxist account. No formal exploitative arrangement exists between labourer and labour thief. Rather, the thief simply takes the products of labour without offering any compensation at all.

The agricultural analogy is suggestive of the connection between labour and property made by John Locke. His account begins from our natural rights to life and self-defence, which originate in self-ownership, and extend to what we do with our selves, that is to say, to our labour and the products thereof:
\begin{quote}
[E]very man [\textit{sic}] has a property in his own person: this no body has any right to but himself. The labour of his body, and the work of his hands, we may say, are properly his. Whatsoever then he removes out of the state that nature has provided, and left it in, he hath mixed his labour with, and joined to it something that is his own, and thereby makes it his property. \cite[][Ch. V, §27]{locke_second_1690}
\end{quote}
In a way, Locke says, our labour adds a part of \emph{ourselves} to that which we work upon. So long as we have a right to use the materials upon which we work---which at this stage in Locke's account is guaranteed, since the resources to be transformed are taken from the commons---we own what we make from them, because we own the labour that produces it, which becomes mixed with the product.\footnote{The way Locke defines a legitimate claim to the use of natural resources is entangled with his racism and personal involvement with colonization and the transatlantic slave trade. I am taking only the part of his account that is concerned with the connection between labour and property, and not the aspects of his account that support taking land from indigenous peoples or provide conditions under which he believed enslavement could be morally justified. For discussion of Locke’s mixed legacy with regard to human rights, see \cite{arneil_john_1996,lewis_locke_2003,mills_locke_2022}} Moreover, since ownership of what we produce derives from our rights to self-ownership and self-defence, we also have the right to protect our labour and its products from theft or exploitation.

Locke is mainly concerned with agricultural or industrial labour and physical property. But his argument has also been influential in the philosophical foundations of \emph{intellectual} property.\footnote{Arguments drawing on the connection between labour and intellectual property rights have been unpopular in the legal literature since the U.S. Supreme Court rejected the ``sweat of the brow'' doctrine \cite{noauthor_feist_1991}. However, I am not arguing for the \emph{legal} merit of a Lockean argument in the current system of intellectual property. Rather, the arguments in this paper are aimed at establishing an ethical case for these rights, as a step towards policy changes. Thanks to Alissa Centivany for discussion of this point.} For instance, writing a century after Locke, Denis Diderot deploys a similar argument as part of his efforts to secure intellectual property rights for authors. Indeed, for Diderot, ownership over the products of one’s mind is more fundamental than ownership over the fruits of physical labour, since, on his view, the mind is the seat of the self:
\begin{quote}
What property can a man [\textit{sic}] own if a work of the mind... if his own thoughts, the feelings of the heart, the most precious part of himself... does not belong to him? ...I repeat, the author is master of his work, or no one in society is master of his property. \cite{diderot_letter_2002}
\end{quote}
On this account, when one creates a piece of creative work—be it a text, painting, song, computer program, whatever—the labour of one’s mind is mixed with the product. And, just as for the products of physical labour, since we own our intellectual labour, we own the creative works that result. An artist owns their art because their creative labour is mixed with the resulting images.\footnote{Some philosophers argue that the labour-based right to property stems from the unpleasantness of labour, since Locke refers to the ``pains'' of labour as deserving compensation. As a result, on this interpretation, intellectual property rights cannot be grounded in labour because creative labour is pleasurable or intrinsically valuable \cite{hughes_philosophy_1988}. I think this objection is a nonstarter for three reasons. First, there are other aspects to Locke’s account of labour and property that are arguably more fundamental—in particular, the derivation of property rights from the right of self-ownership. Since one owns oneself, one must own what one does with oneself, be it painful or otherwise. Second, anyone who knows an artist will know that even if some parts of the creative process are enjoyable, long portions of the work are difficult and unpleasant. It is thus not so simple to claim that creative work does not involve pains. Third, this argument would also apply to physical property. There are plenty of people who take pleasure and find intrinsic value in working hard with their bodies, yet we would not question that they are labouring. We should instead understand Locke’s remarks about the ``pains'' of labour as making a contrast between work and leisure. The point isn't that to labour is to suffer, but rather, that one should be rewarded for choosing productive work over idleness.}  As jurist Frank Easterbrook writes, ``Intellectual property is no less the fruit of one's labor than is physical property'' \cite[][p. 113]{easterbrook_intellectual_1990}.

It is perhaps less clear what our intellectual or creative labour transforms---what is the analogue of unclaimed natural resources in the commons? One set of resources for creative labour is what we may call \emph{creative building-blocks}: language, concepts, ideas, stories, mathematics, culture, artistic processes and principles, and so on. These are components that we recombine in various ways to produce new creative works, and are themselves not the sorts of things over which any one person could assert property rights. Another set of resources for creative labour includes specific creative works that have entered the public domain, such that anyone may freely transform, adapt, and build upon them.

The case against AI image generators built using art scraped from the web can thus be made by way of a Lockean argument. Because artists own the labour that produces their works by transforming creative building-blocks, they thereby own those works. To take those works and use them to create an AI model is analogous to stealing grain from a farmer who grew it, and baking a cake with the stolen ingredient. While additional labour was involved in producing the eventual product, the thief-\textit{cum}-baker has no right to claim ownership of the cake because they had no right to the grain they took to make it. The AI developer has no right, then, to those billions of artworks scraped from the web, and so has no right to use these products of human creative labour in AI development.

\section{Might All Art Be Theft?}
While the Lockean argument against AI art avoids some of the difficulties of the others considered so far, and offers a more general case against text-to-image AI than some of the senses of theft above, it is vulnerable to several objections. Some are rooted in alleged problems with Locke's account itself, such as Robert Nozick's argument that the concept of labour and how it is mixed with materials are unclear \cite[][pp. 174ff.]{nozick_anarchy_1974}. I will set these aside to avoid a digression into Locke scholarship; assume for the sake of argument that an acceptable formal conceptualization of labour is attainable. Another set of concerns may arise regarding how the account so far should inform jurisprudence and policy-making in intellectual property law. These too I set aside, because my interest in this paper is not in the details of policy and judicial rulings but in the ethical principles that should guide such proceedings.

The main objection I will respond to concerns the process of creative labour itself, and the dependence of human artists on the works of others to learn and practise their craft. For, another resource upon which creative labour depends are the ample numbers of works shared by other creative labourers, that is to say, works over which creators still hold a right of ownership. That artists constantly borrow elements of one another's work is no secret. We have seen this already in Kleon's advice; indeed the notion that artists ``steal'' from one another forms the title of his book \cite{kleon_steal_2012}. The Lockean argument so far appears to have the consequence that it would be theft for artists to use one another's works for reference, inspiration, or learning. This conclusion would be unacceptable.

Some version of this objection is central to arguments made by writers critical of how copyright law interfaces with digital media more generally.\footnote{Thanks to an anonymous reviewer for pushing me to respond to this form of the objection.} Jessica Litman observes that the distinction between sharing facts, which cannot be protected by intellectual property rights, and sharing creative works, which can be so protected, seems spurious from a consumer standpoint, and undermines the potential of the internet for facilitating creative projects \cite{litman_sharing_2004}. Lawrence Lessig goes so far as to claim that the restrictions on sharing and remixing media enforced by intellectual property law online is ``silly'' when it is plain that this slows and stifles creativity and innovation by tangling the act of creation in a legalistic mire that is oriented more to protecting the interests of large rights-holder corporations than those of individual artists and scholars \cite{lessig_future_2002,lessig_free_2004}.

Put another way, what is it about the development and use of AI image generators that is relevantly different from a human artist's development and exercise of their creative skills? One might think that if the one constitutes theft, so must the other. Both require training by exposure to existing works, and both draw on existing works to produce new creative products. Moreover, both would be hampered if artists were granted strong property rights that would slow or stop others from drawing on their creations.

A related objection concerns the difference between physical and abstract property. In the argument by analogy presented in the previous section, a farmer's grain is stolen so that the thief may bake something with it. But in cases where the products of creative labour are ``stolen,'' the owner does not seem to lose anything. (Indeed, it was precisely this point that led to the dismissal of the ``heist'' sense of theft above.) While this point alleviates worries about artists potentially stealing from one another, it appears to do so at the expense of the entire Lockean argument. If drawing on the works of other artists is not theft when human artists do it, why would it be theft when AI developers or computer applications do?

We seem to be caught in a dilemma. Since creative labour draws on the products of past creativity, either all (or, at least, much) art is theft, or, no art is theft, including AI art. In the rest of this section, I argue that we can find a way around the horns of this dilemma by considering the ethically relevant differences between the two cases. As I will argue, these differences between how humans and generative AI draw on existing creative work justify us in calling the latter, but not the former, a kind of theft.\footnote{There are at least two other ways of distinguishing human creative labour from AI image generation. Birhane argues that generative AI is a deterministic process, while human creativity is not \cite{birhane_impossibility_2021}. I am doubtful about her approach, because it seems to hold creativity hostage to indeterminism about human agency, bringing it too close to what Boden calls ``pseudo-mysticism'' about creativity \cite[][p. 15]{boden_creative_2004}. Another approach would be to point out differences in cognition between machine learning systems and human minds. For example, Sato and McKinney argue that because generative AI lacks embodiment, it cannot be creative \cite{sato_enactive_2022}. I mention these to set them aside, since they focus on metaphysical issues rather than the ethical distinctions I wish to make.}

One response to these objections could follow the emphasis protestors that place on \emph{consent} to the reuse of one's creative works. Recall the three senses of art theft identified in the previous section. While we might have pre-theoretically thought that to commit theft is to deprive someone of something that they own or have a right to, we can see now that this is not necessarily true. A heist or forgery may deprive an artist of a physical artwork or credit for their work, respectively, but style and labour theft need not deprive the victim of anything. What is in common to each of these senses of theft is, rather, that they are \emph{nonconsensual \emph{uses} of a thing over which the victim has a claim of ownership}. What is wrong about theft, then, is the disrespect for persons constituted by a violation of consent.

Consider these remarks from artist Karla Ortiz, speaking to \textit{The New York Times} about Glaze, an image processing tool that confuses AI algorithms trained on treated images:
\begin{quote}
We're taking our consent back... [AI image generators] have data that doesn't belong to them... That data is my artwork, that's my life. It feels like my identity. \cite{hill_this_2023}
\end{quote}
Without asking for the artist's permission to use the products of their creative labour, we might argue, AI developers are treating the artist merely as a means to their own ends. As Immanuel Kant famously writes, to fail to treat someone as an end in themselves by respecting their own desires, values, and sense of self is to commit an act of disrespect for both the individual person and for humanity in general by undermining their autonomy \cite{kant_groundwork_1785}.

There is a potential issue with leaning too heavily on consent, however. Artists taking inspiration from one another do not generally write to ask permission, even when engaging in imitation in order to improve their skills. So, why would an AI developer owe such consideration? This takes us back to the critique of copyright offered by Lessig and Litman: it seems that, especially but not exclusively online, sharing is the norm when it comes to drawing on other creators' works. Requiring explicit consent to use another visual artist's works sounds like precisely the creativity-stifling copyright régime that these scholars criticize.

One possible response is to point to a difference in purpose between a human artist's borrowing from another artist and an AI developer's or user's borrowing from human artists. Lessig and Litman may be correct that there are permissive norms of sharing amongst human artists. Artists know that sharing their work will in part serve to inspire others in developing their own artistic projects, and this is generally considered good. But until recently, most artists were unaware that data brokers had also been gathering the work that they had shared for the purpose of training AI models. While both uses of shared art lead to the production of new, potentially aesthetically pleasing images, this narrowly consequentialist analysis of the good of art sharing overlooks that the intermediary process is different.

Kai Spiekermann et al. argue that, contrary to what Lessig and Litman suggest, the general norm around taking information from others is actually that it is impermissible \cite{spiekermann_big_2021}.\footnote{I thank an anonymous reviewer for making me aware of similarities between my arguments in this section and the more general arguments about norms for data sharing developed by Spiekermann et al.} Spiekermann et al. point out that copying someone's notes in secret would be impermissible, regardless of the potential good it may do for intellectual, creative, or social projects. We might also refer to theories of privacy based on norms of information sharing \cite[e.g.][]{warren_right_1890,fried_privacy_1968,rachels_why_1975,nissenbaum_privacy_2004,nissenbaum_privacy_2010}. On these accounts, there are established expectations for who is permitted to take which kind of information from us without explicit permission, and who is required to ask. In creative contexts, there are permissive norms of sharing artworks, and established expectations that other human beings will draw upon these for developing their own skills and for inspiration in creating original or derivative works. But there is no such established norm for the mass appropriation of such works for the purposes of creating text-to-image AI models. The lack of such an information norm means that explicit consent is required in the case of AI development, but not in the case of human artistic practice.

To make this clearer, we can compare the sharing of creative works online with the sharing of personal information on social media platforms such as Facebook and LinkedIn. Platforms such as these ostensibly encourage the sharing of these data so that people can make authentic connections between one another and find those whom they know in real life. This exhortation to share leverages norms established offline where people share personal information to various degrees with their friends, coworkers, and colleagues. At the same time, however, the companies that own these platforms make use of these personal data to inform their algorithmic advertising businesses, a purpose which has not always been clearly communicated to social media users. Just as we might be perfectly happy to share personal information for the purposes of connecting with others, but not in order for advertisements to be personalized to our behaviour, so artists might be perfectly happy to share their work for the purposes of inspiring others, but not for the purposes of creating text-to-image AI models. In both instances, data was shared under an established information norm, which AI developers and algorithmic advertisers, among others, have since violated. Explicit consent was ethically required for these uses of data because they go beyond the implicitly accepted norms of the relevant contexts.

It is the breaking of these assumptions---personal information sharing is for connecting with others, art sharing is for inspiring others---that constitutes the violation of autonomy that is at the heart of the accusation of labour theft. This captures the wrong that is associated with the violation of consent stressed by Ortiz, but without the problematic implications raised by Litman and Lessig. This also explains in what sense taking art from the web for training AI is theft: it is a violation of established norms for the use of something that belongs to others.

We can find another ethically relevant difference between AI training and humans taking inspiration from the works of others by turning again to an insight from Locke. In his account of the origin of property rights in the state of nature, Locke puts a proviso on the appropriation of resources from the commons. Continuing from the passage quoted earlier, Locke adds that one can claim resources from the commons through one's labour only if ``there is enough, and as good, left in common for others'' \cite[][Ch. V, §27]{locke_second_1690}. How exactly to interpret this proviso is a point of debate in the literature \cite{varden_lockean_2012, varden_locke_2021}, but the basic idea is one can only take one's fair share from what is common to all. If someone were to seize so much from the commons that no resources, or only poor-quality resources, were left for others, this would be unjust. Hence, there are ethical limits, delineated by the requirements of distributive justice, on how much one may claim with one's labour.

The enough-and-as-good proviso does not apply directly to the problem at hand, since, as mentioned, the transformation of creative works through one's labour does not deprive others of the ability to do the same. Scarcity is a problem only for material resources.\footnote{Though some abuses of intellectual property law as it currently stands can induce scarcity. For example, Getty Images has been criticized (and sued) for charging licensing fees for material that is in the public domain \cite{hiltzik_column_2016}.} However, we can still apply Locke's insight about distributive justice to the use of creative works.

A key component of more recent thinking about distributive justice is that there are more goods than just material resources to be distributed fairly. In John Rawls's famous account, these include social goods such as ``liberty and opportunity, income and wealth, and the bases of self-respect'' \cite[][p. 302]{rawls_theory_1971}. On his view, social goods should be distributed equally, ``unless an unequal distribution of any or all of these goods is to the advantage of the least favored'' [\textit{Id.}]. Such advantages to the least well-off may come about, for example, when those who are the most well-off are required to reinvest their excess wealth into social and economic projects.

We can use these Rawlsian principles to show that text-to-image AI is unjust. The arrival of AI image generators has disrupted the distribution of social goods in creative fields, with no corrective regulation as yet. The promise of generative AI to replace large numbers of creative workers, as mentioned in the introduction, places those who are already the least well off in this area of the economy in an even worse position than before. This deprives artists of opportunities, particularly the opportunity to earn an income from their craft. In the meantime, benefits accrue to actors who are already relatively well-off---namely, large tech companies such as OpenAI, Google, and Microsoft. As Spiekermann et al. put it, ``it is not true that the free use of the global information commons by AI producers is without adverse effects. The adverse effect is the inequality it produces''  \cite[][p. 583]{spiekermann_big_2021}.

Even if some form of compensation scheme were worked out---such as has been suggested by Spiekermann et al., and by Litman and Lessig in different contexts---where artists could be paid royalties for the use of their work to create diffusion models, this unjust distribution of economic opportunities would persist. This is due to what is perhaps the most significant difference between how text-to-image AI and human artists borrow from existing works: the speed and scale at which AI applications operate. Individual human artists draw on a limited number of specific reference works at a time, and produce new works at a relatively slow pace. AI image generators, on the other hand, in a sense draw on all the hundreds of millions to billions of images in their training set at the same time, and, when provided sufficient computing power,\footnote{It is worth reminding ourselves at this point that the required computational power means that the creation and use of generative AI comes at considerable environmental cost \cite{strubell_energy_2019, bender_dangers_2021, luccioni_estimating_2022, crawford_atlas_2021}.} can do so rapidly. AI image generators fail to leave ``enough and as good'' through their borrowing of existing works not by bogarting scarce resources, but by suffocating the competition.

The systematic nature of data scraping also reveals another set of goods whose distribution is rendered unfair by generative AI, namely, what Rawls calls the bases of self-respect. The systematic violation of artists' autonomy through rampant data scraping is a pattern of disrespect that wrongs artists in a capacity central to their humanity. One seeks consent from another when one wishes to ensure that one's intentions align with the values and desires of the other. Not to do so is to impose one's own desires, ignoring the other's autonomy. Explicit consent may not always be required---particularly when there are established conventions to protect one another's autonomy through a general attitude of respect, such as is illustrated in the understanding that human artists may draw upon one another's works, as long as they do not plagiarize. When a person's autonomy is repeatedly and systematically undermined, this in itself undermines a fundamental social basis for self-respect. As social creatures, our self-respect is linked to the respect that we are shown by others. To lose the respect of one's fellows is damaging, if not fatal, to self-respect. AI image generators thus produce another distributive injustice, by broadly undermining the bases of self-respect across creative communities. This, I think, captures part of why Ortiz describes the appropriation of her work in terms of a theft of something deeply personal---to quote her again: ``That data is my artwork, that's my life. It feels like my identity.''

To close this section, it is worth noting that this pattern is a recurring one that we see in data science and machine learning, and it is a pattern with deep and troubling historical roots. Shoshanna Zuboff, following historians of colonization, calls this a \emph{conquest pattern}: a powerful entity appropriates some resource, declares that they have a right to it, extracts value from that resource, and accrues the profits, all of which poses an existential threat to the people from whom that resource was appropriated \cite[][pp. 176ff.]{zuboff_age_2019}. She shows that precisely this playbook has been used by both Spanish conquistadors in central America and Google on the web. On her account, both are instantiations of what Hannah Arendt (following Marx) called capitalism's ``original sin of simple robbery'' \cite[][p. 192]{arendt_origins_1951}, which becomes repeated without limit in the quest for limitless growth, ``continuously laying claim to decision rights over whatever is in its path'' \cite[][p. 139]{zuboff_age_2019}.

In historical colonialism, the conquest pattern took the form of the appropriation of land from indigenous peoples for agriculture and settlement. In what is increasingly called \emph{data colonialism} \cite{couldry_costs_2019, thatcher_data_2016, birhane_algorithmic_2020}, the pattern is to seize data and process it for profit, without regard for the rights and interests of the data subjects involved, all while threatening the bases of a flourishing life for those whose data has been taken. In the case of AI image generators specifically, artists have had their work appropriated \textit{en masse} by more powerful technologists for the creation of generative AI tools that pose an existential threat to artists’ livelihoods. The creation of AI image generators using such methods is thus fundamentally unjust, for it makes those who are already less powerful worse off than they were before.

\section{Conclusion: Just How Much AI is Theft?}
In this paper, I have substantiated an argument made by professional artists against the development and use of AI image generators. I showed how several senses of art theft apply to generative AI, focusing in particular on the notion of labour theft as the central sense because of its dialectical advantages in making the ethical case against text-to-image AI. Using a Lockean account of creative labour, I argued that AI image generators, at least those using diffusion models, involve a large scale and morally objectionable form of theft, rooted in the appropriation of vast numbers of existing artworks. Finally, I showed how the mass appropriation of existing works by AI developers contrasts with the smaller-scale borrowing of existing works by human artists. Unlike human processes of borrowing from existing creative works, the process of creating text-to-image AI produces distributive injustices of both material resources and the bases of self-respect, violates data subjects' autonomy by breaking established information norms, and replicates an exploitative pattern of value extraction continuous with the excesses of colonialism and underregulated capitalism.

If these arguments are right, then there are implications for AI development in general. Many kinds of AI technologies---especially but not exclusively forms of generative AI---require massive datasets in order to be trained. If the use of images (and their descriptions) found on the web for this purpose is morally objectionable, then so is the use of webhosted text, video, audio, and, in general, any form of human-produced creative works obtained without the permission of their creators. If \textsc{dall•e}, Midjourney, and Stable Diffusion are theft, so too are Bard, Bing, ChatGPT, CoPilot, Magenta, and any other form of generative AI built on the appropriation of vast troves of data obtained without consent. Moreover, machine learning models that are not employed in generative systems---such as, perhaps, certain large natural language understanding models---may also involve theft. New approaches to data collection and processing, as well as enforceable regulations codifying the underlying ethical principles, are needed for large AI model development to be morally permissible. Until then, these impressive new technologies do not stand on the shoulders of giants; rather, they parasitize their innards.

%% Bibliography
\bibliographystyle{ACM-Reference-Format}
\bibliography{Citations.bib}

%% Acknowledgements
\begin{acks}
I presented early forms of this paper at the following venues in 2023: the Philosophy, AI, and Society (PAIS) Workshop at Stanford University; the Greater Boston Area Tech Ethics Works-in-Progress Group; a job talk at California Polytechnic State University, San Luis Obispo; at the Philosophy Talk Shop and the Inaugural Conference of Technology Ethics eXchange NorthEast (teXne) at Harvard University; the Congress of the Canadian Philosophical Association at York University; the Discussion Club at the Sage School of Philosophy of Cornell University; and the (Dis)trust and AI Workshop at the University of Western Ontario. I am particularly grateful to Jeff Behrends, Alison Simmons, Jenna Donohue, William Cochran, Kevin Mills, Quinn White, David Collins, Carolyn McLeod, Alissa Centivany, Willow Starr, Z. K. Payne, Henry Ying Kit Kwok, Pat Sewell, and four anonymous reviewers for their feedback. I am also thankful to my partner, Kat Curwin, herself a visual artist, for discussion of this topic and for encouraging me to write this paper. 

Land acknowledgement: Cornell University’s Ithaca campus is located on the traditional homelands of the Gayogoho:no' (the Cayuga Nation). The Gayogoho:no' are members of the Haudenosaunee Confederacy, an alliance of six sovereign Nations with a historic and contemporary presence on this land. The Confederacy precedes the establishment of Cornell University, New York state, and the United States of America. We acknowledge the painful history of Gayogoho:no' dispossession, and honour the ongoing connection of Gayogoho:no' people, past and present, to these lands and waters. More information about the indigenous people of this place and Cornell's connections to colonialism can be found on \href{https://cals.cornell.edu/american-indian-indigenous-studies/about/land-acknowledgment}{the American Indian and Indigenous Studies Program's website}.
\end{acks}

%% LaTeX endmatter
\end{document}